\documentclass[aps,twocolumn,amsmath,amssymb,amsfonts,showpacs]{revtex4}

\usepackage{epsfig}

\newcommand{\beq}{\begin{equation}}
\newcommand{\eeq}{\end{equation}}
\newcommand{\bea}{\begin{eqnarray}}
\newcommand{\eea}{\end{eqnarray}}
\newcommand{\hf} {\frac{1}{2}}
\newcommand{\nn}{\nonumber\\}

\newcommand\eqn[1]{Eq.\,(\ref{#1})}

\newcommand\fig[1]{Fig.\,{\ref{#1}}}
\newcommand\sect[1]{Sect.\,{\ref{#1}}}

\newcommand\app[1]{Appendix~\ref{#1}}

\def\t{\tilde}

\begin{document}

\title{Infrared fixed point in quantum Einstein gravity}

\author{S. Nagy}
\affiliation{Department of Theoretical Physics, University of Debrecen,
P.O. Box 5, H-4010 Debrecen, Hungary}
\affiliation{MTA-DE Research Group in Particle Physics, H-4010 Debrecen P.O.Box 105, Hungary}
\author{J. Krizsan}
\affiliation{Department of Theoretical Physics, University of Debrecen,
P.O. Box 5, H-4010 Debrecen, Hungary}
\author{K. Sailer}
\affiliation{Department of Theoretical Physics, University of Debrecen,
P.O. Box 5, H-4010 Debrecen, Hungary}

\begin{abstract} 
We performed the renormalization group analysis of the quantum Einstein gravity
in the deep infrared regime for different types of extensions of the model.
It is shown that an attractive infrared point exists in the broken symmetric phase
of the model. It is also shown that due to the Gaussian fixed point the IR critical exponent
$\nu$ of the correlation length is 1/2. However, there exists a certain extension of the model
which gives finite correlation length in the broken symmetric phase. It typically
appears in case of models possessing a first order phase transitions as is demonstrated
on the example of the scalar field theory with a Coleman-Weinberg potential.
\end{abstract}
\pacs{11.10.Gh,11.10.Hi,04.60.-m}

\maketitle

\section{Introduction}
The renormalization group (RG) method can create a bridge between theories with different
energy scales \cite{rg}. Generally, one starts with the high-energy ultraviolet
(UV) Langrangian of the model, which describes well the short-range interaction of the elementary
excitations. Furthermore, the UV Lagrangian usually has a very simple form with a small number
of couplings due to a number of symmetry restrictions.
On the other hand, the low-energy infrared (IR) limit of the models is usually
very complicated, where the long-range interactions may induce infinitely many new couplings,
non-localities, or global condensates breaking some symmetries. The IR theory can 
be obtained by the successive elimination of the degrees of freedom with the help of the RG
technique. Similar problems appear in various field-theoric quantum gravity models if one goes
beyond the UV regime \cite{Reuter_rggrav,qg_rev}. The quantum Einstein gravity (QEG)
seems to be non-renormalizable according to perturbative considerations. However it was shown,
that the model has an UV fixed point, which makes QEG renormalizable and provides the asymptotic
safety \cite{Weinberg,Reuter_rggrav,as_rev,as_saf}. The phase space contains a
non-Gaussian UV fixed point (NGFP) and a trivial Gaussian fixed point (GFP). The former is
an UV attractive, i.e. the NGFP attracts the RG trajectories when the UV cutoff is removed to
infinity. It is a focal-type fixed point characterized by complex exponents.
The GFP is sometimes referred an IR one, nevertheless it is a hyperbolic or saddle point type
fixed point, therefore it is rather a crossover type point. The fixed points can be
characterized by universal scaling laws of the corresponding couplings with the appropriate
critical exponents. The scalings, and the exponent in the UV NGFP was extensively studied
so far \cite{Hamber,exps}.

Our goal is to investigate the low-energy IR limit of QEG. Recently it has been shown
that an attractive IR fixed point exists in quantum gravity \cite{Donkin}.
The existence of the attractive IR fixed point was also investigated in \cite{Tetradis,Osborn}
for scalar models, and the IR physics of QEG was also investigated in detail
\cite{ir_qg,Reuter_ir} from several aspects.

It was shown that the broken symmetric phase of the d-dimensional $O(N)$ scalar model
\cite{Nagy_ond} and the 2-dimensional (2d) sine-Gordon (SG) model \cite{Nagy_deg,Nagy_zsg}
also possesses an attractive IR fixed point. In these models one can
define the correlation length $\xi$ as the reciprocal of the scale $k_c$ where
the RG evolution stops enabling us to determine the corresponding critical exponent $\nu$
of $\xi$ in the vicinity of the IR fixed point.
This method enabled us to determine the critical exponents
and the type of the phase transition appearing in scalar models, e.g. it was shown in
the $O(N)$ model that the exponent $\nu$ at the IR fixed point equals
to one that is obtained in the vicinity of the crossover
Wilson-Fisher (WF) fixed point \cite{Nagy_ond}.
Furthermore, the infinite nature of the phase transition in the SG model was also
uncovered by this method \cite{Nagy_deg}, where the IR fixed point has the same
exponent $\nu$ of $\xi$ as the Kosterlitz-Thouless (KT) type fixed point being
a crossover point too.
These examples suggest that the attractive IR fixed point inherits the value of the exponent
$\nu$ from the value taken at the fixed point passed by the RG trajectory previously,
which is typically a hyperbolic type crossover singular point. This comes from the fact that
the broken symmetric phase is
characterized by a condensate built up by the bulk amount of soft modes, and its
global nature appears along the whole flow in the broken symmetric phase.

This argument implies that in quantum gravity the IR fixed point of the broken symmetric
phase may be strongly affected by the GFP. We show numerically that the critical exponent $\nu=1/2$
has the same value at the GFP and at the IR one. Approaching the IR fixed point the scale
parameter $k$ of the RG flows tends to zero, its reciprocal
determines a diverging correlation length which signals a continuous phase transition.

The IR fixed point is situated at
finite dimensionless cosmological constant, so the dimensionful coupling should tend
to zero. We note, however,  that the correlation length does not diverge 
in the neighborhood of the critical point when the phase transition is of a
first order type \cite{ZJ}.
If there is a first order phase transition in the IR limit, then the dimensionful cosmological
constant remains finite, which may provide a possibility to handle the `famous' cosmological
constant problem.

In this article we show that certain extensions of the QEG may show such scaling
behavior in the IR limit, which is typical in models having a first order phase
transition. In order to demonstrate the first order type scaling, we investigate
the scalar model with Coleman-Weinberg (CW) potential which possesses a 
$U(2)\times U(2)$ symmetry \cite{CW,Litim_cw,Berges_cw,Fukushima}.
One can consider the CW model as a prototype of the first order transition.
Certain limit of the model shows an $O(8)$ symmetry giving second
order phase transition and having a crossover type Wilson-Fisher and an IR fixed points. In this
case the degeneracy induced scaling can give us a diverging correlation length. When the general
situation takes place in the CW model, the first order of the phase transition is signaled
by the appearing of a non-trivial minimum of the potential during the RG evolution.
To detect this, one has to determine the evolution of the whole potential \cite{Fukushima}.
It mathematically means that one should solve a partial differential equation (or its
discretized system of ordinary differential equations version), which solution is quite involved,
and can be unstable especially in the broken symmetric phase \cite{Pangon}.
Our method gives an alternative identification for the first order phase transition, which
can be handled numerically in an extremely easy way.
The dynamically generated degeneracy scale appears in the CW model, too.
However, as the RG trajectory approaches the critical point the degeneracy scale
does not tend to zero, but it takes a finite value implying a finite correlation length.
For vanishing value of the coupling $\lambda_2$ of the CW model there is a continuous phase
transition, while for nonzero values of $\lambda_2$ the correlation
length remains finite in the IR limit and a first order phase transition is shown up.
In this manner the identification of the order of the phase transition can be reduced
to a fine tuning of the couplings to their critical values. The other advantage of this
method is that it is not necessary to know where the fixed point is situated in the phase space.

As a rule, various extensions of QEG do not change the type of the scaling in the IR limit,
although it was shown \cite{Reuter_rggrav}, that the IR flows strongly depend on the choice
of the extension. Usually a second order phase transition appears in the IR limit,
but a special extension of QEG is also known in which a first order phase transition
takes place.

It seems to contradict to the fact that a continuous phase transition appears
in the UV regime. Let us note, however, that there are several models which
show up phase transitions of different orders in the UV and in the IR regimes.
For example, in the case of the 2d massive SG model \cite{msg} the UV scaling predicts a KT-type,
infinite order transition, while a complete analysis, which takes into account all the quantum
fluctuations shows that a second order phase transition takes place in the model
\cite{Byrnes,msg}.

The paper is organized as follows. In \sect{sec:ren}
we give a short summary of the RG method. In \sect{sec:first} the RG treatment of the
CW model is presented, and the scaling properties of a first order phase transition are
discussed. The IR fixed point of the phase space of QEG is mapped out and the values of the
critical exponent $\nu$ belonging to the correlation length are determined in
\sect{sec:qeg} for various dimensions. In \sect{sec:ext} we determined the value of
$\nu$ for various types of extensions of QEG, and
put much emphasis on finding such an extension where $\xi$ of QEG
does not diverge suggesting a first order phase transition in its IR limit.
Finally, in \sect{sec:sum} the conclusions are drawn up.

\section{Renormalization}\label{sec:ren}
We consider the RG treatment of the Euclidean QEG.
We note that the Lorentzian form of the model is also investigated thoroughly \cite{Manrique}.
The Wetterich RG evolution equation for the effective average action $\Gamma_k$ reads as
\beq\label{potev}
\partial_t \Gamma_k = \hf\mbox{Tr}\frac{\partial_t R_k}{\Gamma''_k+R_k}
\eeq
with the `RG time' $t=\ln k$, the prime denotes the differentiation with respect to the field
variable, and the trace Tr denotes the integration over all momenta and the summation
over the internal indices. \eqn{potev} is valid for scalar, fermionic or gauge fields, too.
The function $R_k$ plays the role of the IR regulator. We usually use the optimized one
\cite{Litim_opt} of the form
\beq
R_k = (k^2-q^2)\theta(k^2-q^2)
\eeq
which provides fast convergence in the flows and analytic formulae for the evolution equations.
We impose a certain functional ansatz for the effective average action in the local potential
approximation which contains couplings. From the Wetterich RG equation in \eqn{potev} one can
get a coupled system of evolution equations for the dimensionless couplings which is solved
numerically.

The momentum scale $k$ covers the momentum interval from the UV cutoff $\Lambda$ to zero, unless
a singularity occurs at some finite critical momentum scale $k_c$.
During the numerical calculations we set $\Lambda=1$.

\section{First order phase transition}\label{sec:first}
\begin{figure}
\begin{center} 
\epsfig{file=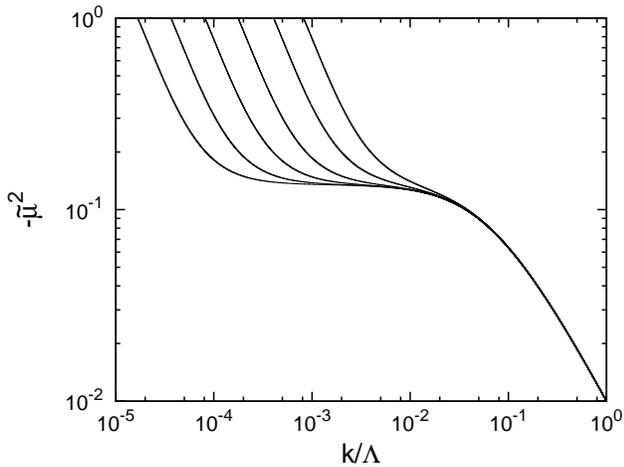,width=6cm,angle=-90}
\caption{\label{fig:sec} The scaling of the coupling $\t \mu$ in the broken symmetric phase
for the CW model is presented.
The curves correspond to evolutions with different initial conditions. The momentum scale
that corresponds to the degeneracy tends to zero as the the reduced temperature approaches zero.
} 
\end{center}
\end{figure}

The CW model exhibits a $U(2)\times U(2)$ symmetry. We summarized the derivation
presented in \cite{Fukushima} in \app{app:cw}. It leads to the following RG equations
for the dimensionless couplings,
\bea\label{eveq}
\partial_t \t\mu^2 &=& -2\t\mu^2-\frac{5\t\lambda_1+18\t\lambda_2}{9\pi^2(1+\t\mu^2)^2},\nn
\partial_t \t\lambda_1 &=& -\t\lambda_1+\frac{8(2\t\lambda_1^2+9\t\lambda_1\t\lambda_2
+54\t\lambda_2^2)}{9\pi^2(1+\t\mu^2)^3},\nn
\partial_t \t\lambda_2 &=& -\t\lambda_2-\frac{4(\t\lambda_1\t\lambda_2+2\t\lambda_2^2)}
{3\pi^2(1+\t\mu^2)^3},
\eea
where a second order phase transition appears when $\lambda_2$ does not evolve. Then
in the broken symmetric phase the correlation length diverges as the reduced
temperature $t$ tends to zero. It is defined as $t\sim\t\lambda_{1\Lambda}^*
-\t\lambda_{1\Lambda}$, where the lower index $\Lambda$ refers to the value at the UV scale
$\Lambda$ and the upper index $^*$ denotes the corresponding UV critical value.
We numerically determined the divergence of $\xi$ and
show the obtained results in \fig{fig:sec}.
The RG equations in \eqn{eveq} become degenerate, which means that the expression
of the form $1+\t \mu^2$ in the denominator tends to zero.
As the temperature approaches its critical value, the momentum scale of the degeneracy tends
to zero and the correlation length diverges signaling a continuous phase transition in the model.

\begin{figure}
\begin{center} 
\epsfig{file=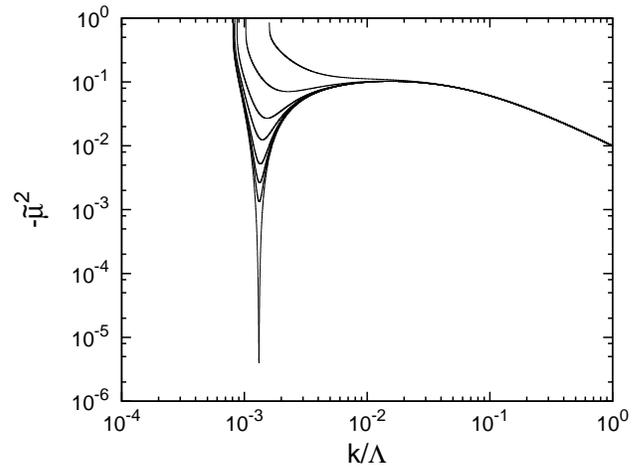,width=6cm,angle=-90}
\caption{\label{fig:fir} The scaling of the coupling $\t \mu$ in the broken symmetric phase
is shown. The curves correspond to evolutions with different initial conditions.
The degeneracy momentum scale tends to a finite value as  the reduced temperature 
approaches zero.
} 
\end{center}
\end{figure}

\begin{figure}
\begin{center} 
\epsfig{file=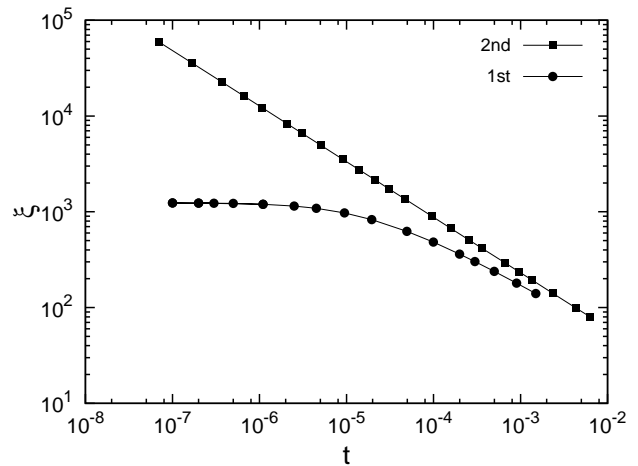,width=6cm,angle=-90}
\caption{\label{fig:cwcorr} The dependence of the
correlation length on the reduced temperature is presented. For the second order
phase transition $\xi$ diverges as $t\to 0$, while in the case of the first order phase
transition it tends to a constant value for $t\to 0$.
} 
\end{center}
\end{figure}

When the coupling $\lambda_2$ evolves then the order of the phase transition changes
from a second to a first order one. This can also be catched by the flow
of $\t \mu^2$ in \fig{fig:fir}.
As the temperature approaches its critical value, the negative coupling $-\t \mu^2$ does
not grow monotonically but it tries to change its sign. The change appears when $t=0$.
The degeneracy also appears but its corresponding momentum scale $k$ does not tend to zero
as $t\to 0$ but the flows stop at the same critical value of $k$. We plotted the
scaling of the correlation length $\xi$ as the function of the reduced temperature $t$
in \fig{fig:cwcorr}.
The results show that $\xi$ diverges as a power law $\xi\sim t^{-1/2}$ in the case
of the second order phase transition, while $\xi$ tends to a constant value for a
first order phase transition.

Thus, the degeneracy induced IR scaling of the correlation length
can account for the type of the phase transition in a very simple way.

\section{Quantum Einstein gravity}\label{sec:qeg}

\begin{figure}
\begin{center} 
\epsfig{file=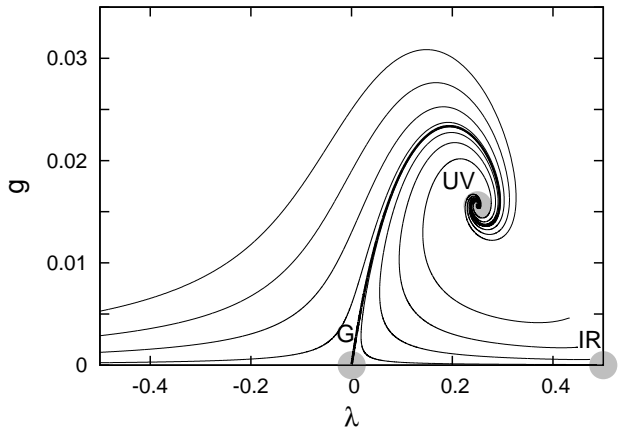,width=6cm,angle=-90}
\caption{\label{fig:phase} The phase structure of quantum Einstein gravity.
There is an UV non-Gaussian, a crossover Gaussian and an IR fixed point. The thick
line represents the separatrix.
} 
\end{center}
\end{figure}

The QEG effective action is defined as
\beq\label{eaqeg}
\Gamma_k = \frac1{16\pi G_k}\int d^d x\sqrt{\mbox{det}g_{\mu\nu}}(-R+2\Lambda_k)
\eeq
with the metrics $g_{\mu\nu}$, the Ricci scalar $R$ and with the couplings, i.e. with
the dimensionful Newton constant $G_k$ and the
cosmological constant $\Lambda_k$. \eqn{potev} provides the evolution equations
for the dimensionless couplings $g =k^{d-2} G_k$ and $\lambda = k^{-2}\Lambda_k$.
The phase space of the QEG is spanned by the evolving parameters $g$ and $\lambda$. By using
the optimized RG regulator one obtains the following evolution equations \cite{Litim_PoS},
\bea\label{gleveq}
\partial_t \lambda &=& -2\lambda+\frac{g}2 d(d+2)(d+5)\nn
&&-d(d+2)\frac{g}2\frac{(d-1)g+\frac{1-4\lambda(1-1/d)}{d-2}}{g-g_b},\nn
\partial_t g &=& (d-2)g+\frac{(d+2)g^2}{g-g_b}
\eea
with
\beq
g_b = \frac{(1-2\lambda)^2}{2(d-2)}.
\eeq
One can introduce the gravitation anomalous dimension
\beq
\eta = \frac{(d+2)g}{g-g_b}.
\eeq
In case of $d=4$ the RG flow equations in \eqn{gleveq} possess two fixed points,
which can be analytically identified, these  are shown in \fig{fig:phase} (the points UV and G).
There is a UV non-trivial fixed point (NGFP) at finite values of $g^*=1/64$ and
$\lambda^*=1/4$. Here the eigenvalues of the linearized RG transformation are complex, 
and the RG trajectories run away with decreasing scale parameter $k$  from the UV fixed
point as if it were a repulsive focal point,
or otherwise one can say, that one has to do with a UV attractive fixed point.
Furthermore, there is a trivial, saddle point-like Gaussian fixed point at the origin of the
phase space $g^*=0$ and $\lambda^*=0$. There the linearized
RG transformation has the eigenvalues
$s_1=-2$ and $s_2=2$. The negative reciprocal of the negative eigenvalue gives
the exponent $\nu=1/2$. The GFP can be considered as a crossover (CO) fixed point 
between the UV and the IR scaling regimes.
The particular trajectory running from the UV fixed point into the 
Gaussian one, the separatrix splits the phase space into two regions according to
the two phases of the model (see \fig{fig:phase}).
The trajectories approaching the separatrix from the left give negative
values for the cosmological constant and vanishing Newton constant in the IR limit,
the corresponding phase can be called the strong-coupling or symmetric one
\cite{Reuter_rggrav,Polonyi_qg}. There the scale $k$ has a well-defined limit $k\to 0$.
Other trajectories getting around the separatrix from the right give
large positive values of $\lambda$ and small Newton constant if the RG flows tend
to the IR regime. This phase can be called the weak-coupling or broken symmetric phase.
There are several models known which have similar phase structure. The non-linear sigma
model for $d>2$ exhibits a symmetric and a broken symmetric phase,
furthermore the model has a NGFP in the UV \cite{nlsm}. The 2d SG model
also has two phases and an UV NGFP \cite{Nagy_zsg}. There the UV limit leads to singularity
of the RG equations. A similar behavior can be found in the IR behavior of the
broken symmetric phase. The latter seems to be typical, since there are many examples where
the RG flows tend to a singular regime in the broken symmetric phase of the model.
We argue that this degeneracy induced singularity is a consequence of the appearing IR
fixed point \cite{Nagy_zsg,Nagy_ond,Nagy_deg,Nagy_per}. We show that the IR fixed point
also exists in QEG in the positive cosmological constant regime.

The RG flow shown in  \fig{fig:phase} indicates that there is an attractive IR fixed point
in the broken symmetric phase of the model at $g^*=0$ and $\lambda^*=1/2$. Although
the IR fixed point does not seem to satisfy the Eqs. in (\ref{gleveq}), since they give
expressions like $0/0$, a suitable reparametrization of the couplings enables one to uncover
this fixed point \cite{Tetradis}. The singularity of the RG flows
seems to be a numerical artifact of the strong truncation of the functional ansatz for the
effective action in \eqn{eaqeg}, but it was shown for other models in \cite{Nagy_ond} that
such a singularity possesses a specific scaling behavior induced by the IR fixed point, therefore
it has significant physical importance. This takes place in QEG, too.
We introduce the new couplings according to $\chi=1-2\lambda$,
$\omega=4g-(1-2\lambda)^2$ and the new 'RG time' $\partial_\tau=\omega \partial_t$. We note
that the idea of redefining the 'RG time' was already used in QEG \cite{Codello}.
Then in the case of $d=4$ the evolution equations can be written as
\bea
\partial_\tau\chi &=& -4\omega+2\chi\omega(8+21\chi)\nn
&&+24\omega^2+6\chi^2(3\chi(\chi+1)-1),\nn
\partial_\tau\omega &=& 8\omega^2(1-6\chi)-2\chi(42\chi^2+9\chi-4)\nn
&&-6\chi^3(\chi(6\chi+5)-2).
\eea
These flow equations have three different fixed points. The Gaussian fixed point appears
at $\omega^*_G=-1$ and $\chi^*_G=1$ with hyperbolic nature. The focal-type UV NGFP can be
identified by the fixed point $\omega^*_{UV}=-3/16$ and $\chi^*_{UV}=1/2$. However
a new fixed point appears at $\omega^*_{IR}=0$ and $\chi^*_{IR}=0$
that can be identified with the IR fixed point and corresponds to
$g^*_{IR}=0$ and $\lambda^*_{IR}=1/2$. The IR fixed point shows very slight,
marginal attractivity. The above reparametrization of the original couplings
could successfully uncover the IR physics of the d-dimensional $O(N)$ model \cite{Nagy_ond},
even in the 2d $O(2)$ case, where an essential scaling signalling an infinite order
phase transition appears, furthermore this simple trick was capable of describing
the phase structure and the order of the phase transition in periodic models
\cite{Nagy_zsg,Nagy_deg,Nagy_per}.

\begin{figure}
\begin{center} 
\epsfig{file=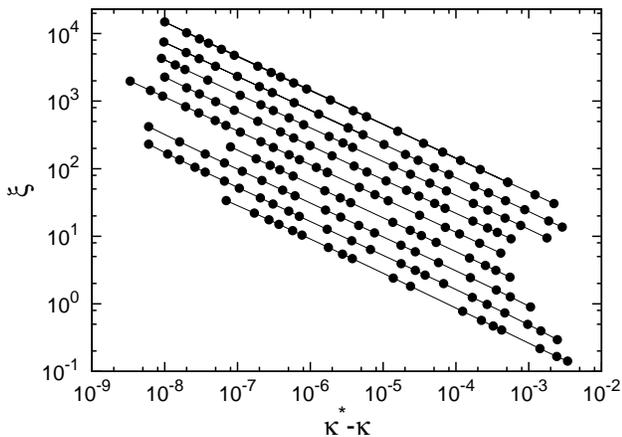,width=6cm,angle=-90}
\caption{\label{fig:dcorr} The scaling of the correlation length, giving $\nu=1/2$
for the exponent. For simplicity we fixed the UV value of $g_\Lambda$.
} 
\end{center}
\end{figure}

The gravitation anomalous dimension $\eta$ shows singular behavior in the IR limit of the
weak coupling phase as $g\to g_b$ at a finite value of $k_c$. It shows that there are no
modes in the theory which can be characterized by the scale where $k<k_c$. However, if
we introduced the RG scale $k'=k_c-k$, then the IR scaling behavior would become visible just
as in \cite{Nagy_ond}. The new RG scale does not cause any trouble, since it is just a
bookkeeping device to enumerate the modes to be integrated out. The scale $k_c$
can be identified by the reciprocal of the correlation length, i.e. $\xi\sim 1/k_c$.
The singularity appears due to the bulk amount of soft modes close to the instability
of the RG flows \cite{Alexandre,Polonyi_qg,Lauscher_cond}.
As the UV value of the couplings $g$ and $\lambda$ approach their critical value, the flow
reaches smaller and smaller value of $k_c$, giving diverging $\xi$. We determined
the values of $\xi$ as the function of $\kappa=g_\Lambda\lambda_\Lambda$ measured from its UV
critical value $\kappa^*$, and plotted it in \fig{fig:dcorr}.
The results show that $\xi$ scales in the IR scaling regime according to a power law function
\beq
\xi \sim (\kappa-\kappa^*)^{-\nu},
\eeq
with exponent $\nu=1/2$. We considered the problem with varying dimension $d$, and
plotted the scaling of $\xi$ for $d=4\dots 11$. We obtained that the IR scaling
behavior and the value of $\nu$ is insensitive to the dimension.

The value of $\nu$ obtained in the IR is in contrast to the UV result $\nu=1/3$ \cite{Hamber}.
Furthermore the UV exponent $\nu$ can also be related to the imaginary part
$\theta''$ of the scaling exponent in the UV fixed point as $\nu=1/\theta''$ \cite{exps}.

The apparent discrepancy can be easily resolved with the following argument.
The fixed points for QEG determine well separated scaling regimes in the phase space, 
and the critical exponents should be calculated in each regions one by one, therefore
it is not surprising that we obtain different values for $\nu$.
Around the UV NGFP  one obtains $\nu_{UV}=1/3$, around the crossover GFP
one has $\nu_{CO}=1/2$, which shows the mean field nature of the model there.
In the IR regime we have $\nu_{IR}=1/2$.
The latter two values coincide. It is not evident, since other exponents in the CO and in the
IR regime may differ. The value of $\eta$ is approximately zero in the CO, while in the
IR it diverges. Similar behavior appeared in the 2d SG model \cite{Nagy_deg}.
The coincidence of $\nu$ in the CO and in the IR may come from the fact that the condensate
can affect the CO regime due to its global property.

There is an obvious similarity between the phase structures of the QEG model and 
scalar models. They usually contain a spontaneously broken and a symmetric phase and it seems
that the broken phase contains an IR fixed point.
In the $O(N)$ model the WF fixed point, in the 2d SG model the Kosterlitz-Thouless (KT) fixed
point plays the role of the CO one. These are the analogues of the crossover Gaussian
fixed point in the QEG model. The $O(N)$ model, the 2d SG model, and the QEG bears an
IR fixed point and the IR value of the exponent $\nu$ of $\xi$ equals the the value
obtained in the corresponding CO fixed point.

The coincidence may suggest that the IR fixed point analysis is
unnecessary, since the CO fixed point has all the information on $\nu$. We note however, that
there are models where there are no CO fixed points but the IR one exists, which provides
the only possibility to get the value of $\nu$ and get the order of the phase transition
\cite{Nagy_per}.
On the other hand it is much easier to calculate numerically the value of $\nu$
in the IR than in the CO, since the former needs a fine tuning of (usually) one coupling
without any knowledge of the fixed point, while
in the latter case we have to find the exact place of the fixed point which is a difficult
mathematical task.

\section{Extensions of QEG}\label{sec:ext}

\begin{figure}
\begin{center}
\epsfig{file=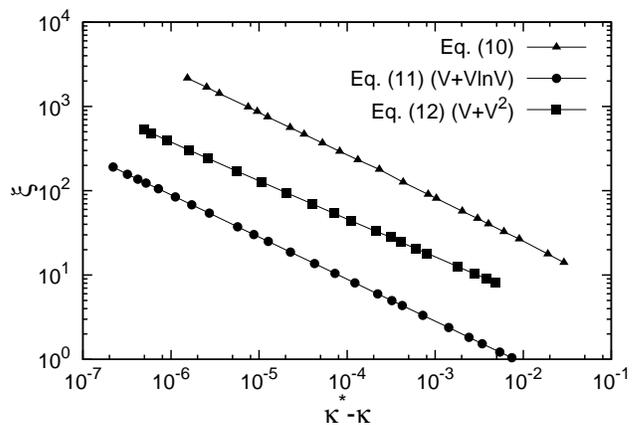,width=6cm,angle=-90}
\caption{\label{fig:corr} The scaling of the correlation length for various extensions of QEG.
The scaling shows second order phase transition with exponent $\nu=1/2$.
}
\end{center}
\end{figure}

Using different forms for the IR cutoff to suppress the low momentum modes of
the graviton and the ghost fields \cite{rg_sch_qg,ext_gh} one can get evolution
equation of qualitatively same form including equations which can be also singular
at certain values of the cosmological constant. Further extension of QEG can be obtained
via e.g. including matter field \cite{ext_matter} or higher order terms in the
curvature scalar \cite{hd_qg}.
Considering one possibility of the IR regularization \cite{Reuter_rggrav,Codello} one obtains
e.g.
\bea
\partial_t \lambda &=& -2\lambda + \frac{g}{6\pi}\frac{3-4\lambda-12\lambda^2-56\lambda^3
+\frac{107-20\lambda}{12\pi}g}{(1-2\lambda)^2-\frac{1-10\lambda}{12\pi}g},\nn
\partial_t g &=& 2g - \frac{g^2}{3\pi}\frac{11-18\lambda+28\lambda^2}
{(1-2\lambda)^2-\frac{1-10\lambda}{12\pi}g}
\eea
for the couplings. Our numerical results showed that the IR scaling is similar to the
previous ones obtained from Eqs in (\ref{eveq}). We got a similar second-order phase transition,
with the same exponent $\nu=1/2$, as is shown in \fig{fig:corr} (indicated by triangles).

Another possible extension of QEG can be achieved by introducing terms to its effective action
containing the functions of the Euclidean spacetime volume
$V=\int d^dx\sqrt{\mbox{det}g_{\mu\nu}}$.
These terms introduce new couplings, which increases the dimension of the phase space.

When the new term has the form of $V+V\ln V$, with its coupling $u$, then the extended
evolution equations become
\bea
\partial_t \lambda &=& -2\lambda+\frac{g}{\pi}
\left(\frac5{1-2\lambda-u}-4\right),\nn
\partial_t g &=& (d-2)g,\nn
\partial_t u &=& -2 u+\frac{10}{\pi}\frac{gu}{(1-2\lambda-u)^2}
\eea
with the choice of the optimized cutoff. If we use the term of the form
$V+V^2$ with the corresponding coupling $w$ we obtain
\bea
\partial_t \lambda &=& -2\lambda+\frac{g}{\pi}\left(\frac5{1-2\lambda}-4\right)
+\frac{32gw}{1-2\lambda},\nn
\partial_t g &=& (d-2)g,\nn
\partial_t w &=& -6 w+\frac{5gw}{\pi(1-2\lambda)^2}+\frac{1024\pi gw^2}{(1-2\lambda)^3}.
\eea
The scaling of $\xi$ is shown in \fig{fig:corr} for different types of extensions,
too. The results show, that the extended models
exhibit a second order phase transition with the same critical exponent
$\nu=1/2$. This IR scaling is also driven by the GFP similarly to the previous results.

One can learn from these examples that if the hyperbolic-type GFP exists then it results in
a continuous phase transition in the IR limit with a diverging correlation length.
The extensions of QEG can introduce new couplings and increase the dimension of the
phase space. There can appear further relevant and irrelevant couplings which
can modify the value of the exponent $\nu$ from its mean field value ($\nu=1/2$) to its
physical one just as in the case of the $O(N)$ models \cite{Nagy_ond}. Other extensions
may give new fixed points or shift  the GFP from the origin \cite{Eichhorn} which
might strongly affect the IR scaling behavior.

However, if we choose the $V+\sqrt{V}$ as an additive term to the effective action with
the coupling $v$ then the evolution equations are \cite{Reuter_ir}
\bea\label{1steveq}
\partial_t  \lambda &=& -2\lambda+8\pi g\left[-\frac1{2\pi^2}+\frac{1-2\lambda}{v^2}\right],\nn
\partial_t g &=& (d-2)g,\nn
\partial_t v &=& \frac{8\pi g}{v}.
\eea
We note that the equations loose their validity when $v\to 0$. It is apparent
that the GFP does not exist in this type of extension. Since the continuous-type
phase transition is strongly related to the existing hyperbolic-type GFP, we do not
necessarily expect a second order phase transition in the IR with diverging
correlation length $\xi$, in principle any type of phase transition might appear \cite{Nagy_per}.

\begin{figure}
\begin{center}
\epsfig{file=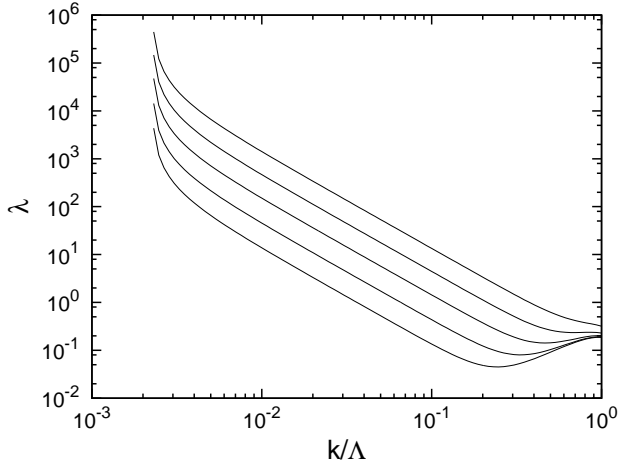,width=6cm,angle=-90}
\caption{\label{fig:vsqv} The flow of the cosmological constant for the
extension $V+\sqrt{V}$ for various initial values of $\lambda_\Lambda$,
$g_\Lambda=1$ and $v_\Lambda\approx 5$. There is a finite momentum scale $k_c$ where the
evolution stops.
} 
\end{center}
\end{figure}

There exists an analytic solution of the flow equation in Eqs. (\ref{1steveq}) which reads as
\beq
\lambda = -\frac{8G^2(2k^6{-}3k^4\Lambda^2{+}\Lambda^6)+6G(\Lambda^4{-}k^4)
-3\lambda v_\Lambda^2}{3(v^2_\Lambda-8\pi G(\Lambda^2-k^2))}.
\eeq
The flows of the cosmological constant are plotted in \fig{fig:vsqv}.
It shows that a dynamical momentum scale $k_c$ appears during the flow, where
the evolution of $\lambda$ becomes singular and stops. It means that
in the broken symmetric phase one cannot go to an arbitrary small scale $k$, implying
that the correlation length does not diverge. According to the example of the CW model
one suggests that
there is a first order phase transition between the weak- and the strong-coupling phases.
This extension of QEG seems to make the originally second order phase
transition into a first order one, due to the vanishing GFP there.

This result also implies that the dimensionful cosmological constant remains finite
in the IR, which is needed to explain the cosmological constant problem.

\section{Summary}\label{sec:sum}

We investigated the infrared behavior of the quantum Einstein gravity with the help
of the renormalization group method. We showed that there exists an attractive IR fixed point
in the broken symmetric phase of the model. It is generally characterized by a diverging
correlation length with the corresponding exponent $\nu=1/2$, showing the mean field nature of the
continuous phase transition which is inherited by the hyperbolic crossover Gaussian fixed point.

This property holds for any dimension, but not for any type of extension for the
effective action. We showed that there exists such type of extension, where the IR degeneracy
defined correlation length does not diverge implying a first order phase transition
in the IR limit. It seems to be the consequence of the disappearing GFP in this extension.
The mechanism is demonstrated via the Coleman-Weinberg model, which is a
typical example for first order transitions.

The appearance of the first order phase transition gives a finite but small value
of IR dimensionful cosmological constant, which may suggest a possible explanation of the
cosmological constant problem.

\section*{Acknowledgments}

The work is supported by the project T\'AMOP 4.2.1./B-09/1/KONV-2010-0007. The project is
implemented through the New Hungary Development Plan co-financed by the European Social
Fund, and the European Regional Development Fund.

\appendix

\section{The Coleman-Weinberg model}\label{app:cw}

The Lagrangian of the $U(2)\times U(2)$ symmetric scalar model is given by \cite{Fukushima}
\bea
{\cal L} &=& \hf \mbox{tr}[\partial_\mu\Phi\partial^\mu\Phi^\dagger]+\hf\mu^2[\Phi\Phi^\dagger]\nn
&&-g_1(\mbox{tr}[\Phi\Phi^\dagger])^2-g_2\mbox{tr}[\Phi\Phi^\dagger\Phi\Phi^\dagger],
\eea
with the dimensionful couplings $\mu^2$, $g_1$, $g_2$ and the $2\times 2$ matrix field variable
which is parametrized as
\beq
\Phi=\Sigma+i\Pi=\sum_\mu t_\mu(\sigma_\mu+i\pi_\mu),
\eeq
where $t_\mu$ are the $U(2)$ generators, $\mbox{tr}[t_\mu t_\nu]=\delta_{\mu\nu}$, and
\bea
\Sigma &=& \begin{pmatrix} \frac1{\sqrt{2}}(a^0+\sigma) & a^+ \cr
a^- & \frac1{\sqrt{2}}(-a^0+\sigma) \end{pmatrix} ,\nn
\Pi &=& \begin{pmatrix} \frac1{\sqrt{2}}(\pi^0+\eta) & \pi^+ \cr
\pi^- & \frac1{\sqrt{2}}(-\pi^0+\eta) \end{pmatrix},
\eea
where $a^0=\sigma_3$, $a^\pm=(a^1\mp ia^2)/\sqrt{2}=(\sigma_1\mp i\sigma_2)/\sqrt{2}$,
$\sigma=\sigma_0$, $\pi^0=\pi_3$, $\pi^\pm=(\pi_1\mp i\pi_2)/\sqrt{2}$ and $\eta=\pi_0$.
The potential of the model becomes
\beq
U = V + W\xi,
\eeq
where
\bea
V &=& \hf \mu^2 \phi^2+\frac1{24}\lambda_1\phi^2,\nn
W &=& \lambda_2,
\eea
with
\bea
\varphi &=& \sigma^2+\vec\pi^2+\eta^2+\vec a^2,\nn
\xi &=& (\sigma^2+\vec \pi^2)(\eta^2+a^2)-(\sigma\eta-\vec\pi\vec a)^2,\nn
\lambda_1 &=& 24\left(g_1+\hf g_2\right),\nn
\lambda_2 &=& 2g_2.
\eea
By using the \eqn{potev} one arrives at the functional equations in dimension $d=3$,
\beq
\partial_k V = \frac{k^4}{6\pi^2}\left(\frac1{E^2_\sigma}+\frac4{E^2_{\pi\eta}}+\frac3{E^2_a}
\right),
\eeq
with
\bea
E^2_\sigma &=& k^2+2V'+4V''\varphi,\nn
E^2_{\pi\eta} &=& k^2+2V',\nn
E^2_a &=& k^2+2V'+2W\varphi
\eea
and for $W$,
\bea
\partial_k W &=& -\frac{k^4}{6\pi^2}\biggl\{\frac{8W'+4W\varphi^{-1}}{E^4_{\pi\eta}}
-\frac{8W^2}{E^6_{\pi\eta}}\nn
&&+\biggl[-4V''\varphi^{-1}+4W''\varphi+10 W'+2W\varphi^{-1}\nn
&&+\frac{8(V''+W+W'\varphi)^2}{2V''-W}\varphi^{-1}\biggr]\frac1{E^4_\sigma}\nn
&&+\biggl[4V''\varphi^{-1}+14W'-6W\varphi^{-1}\nn
&&-\frac{8(V''+W+W'\varphi)^2}{2V''-W}\varphi^{-1}\biggr]\frac1{E^4_a}\biggr\},
\eea
from which the flow equations for the couplings can be derived.

\end{document}